\documentclass[a4paper]{aa}
\usepackage{graphicx}
\def\spose#1{\hbox to 0pt{#1\hss}}
\def\lta{\mathrel{\spose{\lower 3pt\hbox{$\mathchar"218$}}
     \raise 2.0pt\hbox{$\mathchar"13C$}}}
\def\gta{\mathrel{\spose{\lower 3pt\hbox{$\mathchar"218$}}
     \raise 2.0pt\hbox{$\mathchar"13E$}}}

\newcommand{\etal}{et al. }

\begin{document}

   \thesaurus{ 
        10.07.2;
        10.15.1;
        11.19.4;
        11.05.2;
        }
   \title{Evolutionary Synthesis of Simple Stellar Populations}

   \subtitle{Colours and Indices}

   \author{O. M. Kurth, U. Fritze -- v. Alvensleben and K. J. Fricke}

   \offprints{O. Kurth}

   \institute{Universit\"atssternwarte G\"ottingen\\
              Geismarlandstr. 11, 37083 G\"ottingen, Germany\\
              email: okurth@uni-sw.gwdg.de
             }
   \authorrunning{Kurth et al.}
   \titlerunning{Evolutionary Synthesis of SSPs}

\date{Submitted March 13, 1998}

\maketitle

\begin{abstract}
We construct evolutionary synthesis models for simple stellar
populations using the evolutionary tracks from the Padova group (1993,
1994), theoretical colour calibrations from Lejeune et
al. (\cite{lejeune}, \cite{lejeune1}) and fit functions for stellar
atmospheric indices from Worthey et al. (\cite{worthey}).

A Monte-Carlo technique allows us to obtain a smooth time evolution of
 both broad band colours in UBVRIK and a series of stellar absorption
features for Single Burst Stellar Populations ({\bf SSPs}). We present
colours and indices for SSPs with ages from $1 \cdot 10^9$ yrs to $1.6
\cdot 10^{10}$ yrs and metallicities $[M/H]$ = -2.3, -1.7, -0.7,
-0.4, 0.0 and 0.4.

Model colours and indices at an age of about a Hubble time are
in good agreement with observed colours and indices of the Galactic
and M 31 GCs.

\end{abstract}

\section{Introduction}
Colour distributions of Globular Cluster ({\bf GC}) systems are
observed for a large number of early-type galaxies (E, S0, dE, cD)
using ground-based Washington or HST broad band photometry. In most
cases double-peak or broad/mul\-ti-peak colour distributions are seen
(e.g. Zepf \& Ashman \cite{zepf}, Elson \& Santiago \cite{elson},
Kissler-Patig \etal \cite{kissler}).

If the different colour subpopulations of GCs are formed in different
events then they may contain clues to the formation history of their parent
galaxies.  For example a two-peak colour distribution may result, if in addition to
a primary initial collapse population of GCs, a secondary population of
GCs were formed either in a merger-induced starburst (Schweizer
\cite{schweizer}, Ashman \& Zepf \cite{ashman}, Fritze -- v.
Alvensleben \& Gerhard \cite{fritze1}, Fritze -- v. Alvensleben \&
Burkert \cite{fritze2}) or else in some distinct secondary phase of
cluster formation within the original galaxy (Forbes \etal
\cite{forbes}). Likewise the broad or multi-peaked colour distribution often
observed in GC systems around cD galaxies may point to a
series of GC formation events during the hierarchical assembly of the
parent galaxy or to some protracted GC formation or accretion
mechanism.

A well-known difficulty with the interpretation of colour
distributions is the degeneracy of colours with respect to age and
metallicity.  While for Washington photometry there are well
established and reliable calibrations of colours in terms of
metallicity, the situation with HST broad band observations of GC
systems is less clear.  A better understanding of the formation of
composite GC systems would be possible if separate age and metallicity
distributions could be disentangled from an observed colour
distribution.

A second issue concerns the interpretation of colours for young star
cluster systems detected with HST in many interacting and starburst
galaxies.  The question is, if these YSC systems -- at least some
fraction of them -- are the progenitors of GC systems. In an attempt
to answer this question star clusters are being imaged with HST in an
age sequence of interacting galaxies -- from early stages of
interaction through merger remnants up to E/S0s (eg. Schweizer \etal
\cite{schweizer1}, Whitmore \etal \cite{whitmore}, Miller \etal
\cite{miller}).
With 10 m class telescopes, spectroscopy of the brighter members of
young star cluster populations is becoming possible (Kissler-Patig
\etal \cite{kissler1}, Brodie \etal \cite{brodie}, but see also 4 -- 5
m class spectra by Schweizer \& Seitzer \cite{schweizer} or Zepf \etal
\cite{zepf1}).  Spectroscopy will only be possible for a subsample of
YSCs. Thus the determination of ages and metallicities from broad band
colors will still be necessary.

It is thus desirable to study the evolution of broad band colours and
absorption indices for single burst stellar populations of various
metallicities using the most recent and complete stellar evolutionary
tracks as well as careful colour and index calibrations.This allows
one to obtain theoretical calibrations of broad band colours and
indices in terms of metallicity over the full range of ages under
investigation, i.e. from $10^7$ yr to a Hubble time.
 
Since theoretical calculations for the evolution of stars are only
available for a discrete grid of masses, some means for obtaining a
smooth evolution of the composite population is needed. Applying the
tracks as they are would create discontinuities because all stars of a
given mass would reach the giant branch at the same time, dominating
the integrated light until they die. This effect is large for
populations with stars that have about the same age. The effect also
increases with age of the whole population, since the differences in
both the lifetimes and luminosities between the main
sequence and the later stages increase with decreasing mass.

For this work, we use the \emph{Monte Carlo} method to bypass this
problem while still avoiding the interpolation of tracks with its
accompaining danger of creating artificial states. This is described
in detail in section \ref{sec_nummethod}.

The star formation history of any stellar system can be described by a
superposition of SSP models of different ages and metallicities. An
example of this is given by Cellone \& Forte et al (\cite{cellone}) in
their study of Low Surface Brightness galaxies or Contardo et
al. (\cite{contardo}) who investigate the formation and evolution of
galaxies in a cosmological scenario.

\section{Model description}
\subsection{Input physics}
We use the evolutionary tracks of the Padova group (Bressan et al.
\cite{bressan}, Fagotto et al. \cite{fagotto1}, \cite{fagotto2}, 
\cite{fagotto3}, Hereafter referred to as the \emph{Padova tracks}). 
The Padova group gives effective temperatures and luminosities as a
function of time for many masses ($ 0.6 - 120 M_\odot$)
and a metallicity range from $Z = 0.0001$ to $Z= 0.05$. Their tracks
include all stages in stellar evolution from the zero age main
sequence to the tip of the RGB and from the zero age horizontal branch
to the tip of the EAGB. For the stellar mass loss we use the method
described in Bressan et al.(\cite{bressan}) with Reimers law.

For lower masses ($0.08 M_\odot$ to $0.5 M_\odot$) we use the
calculations from Chabrier and Baraffe (\cite{chabrier}). They based
their calculation on a new description of the interiour of low mass
objects and use non-grey atmospheres. Since their grid in metallicity
does not match the grid of the Padova group we have linearly
interpolated the values for $[M/H] = -1.7$, $[M/H] = -0.7$ and $[M/H]
= -0.4$ while for $[M/H] = -2.3$ and $[M/H] = 0.4$ we used the
calculations for $[M/H] = -2.0$ and $[M/H] = 0$, respectively.  The
contributions of these low mass stars to the integrated light is very
low, therefore the error should be small.

To obtain synthetic colours, the conversion from theoretical quantities
to observable quantities is very important. The evolutionary tracks
for stars give effective temperatures, bolometric luminosities and
gravities at the surface of stars of different masses as a function of
time.  These values have to be converted to colours in the various
bands and to atmospheric indices of the evolutionary states.

For the fluxes and colours we use the theoretical library of model
atmosphere spectra for various metallicities of Lejeune et al.
(\cite{lejeune}, \cite{lejeune1}). Lejeune et al. have assembled a coherent library of
synthetic stellar atmosphere calculations from Kurucz (see eg. Kurucz
\cite{kurucz}), Fluks et al. (\cite{fluks}) and Bessell et
al. (\cite{bessell1}, \cite{bessell2}).  The library has an effective
temperature range from $T_{eff} = 2000 K$ to $50000 K$ and covers a
broad range of metallicities. To cope with discrepancies between
colours derived from model atmosphere spectra and observed colours,
they correct (i.e. \emph{bend}) the model spectra to give agreement
with observed colours for U through K. To fit their metallicity grid
to that of the stellar evolutionary tracks, colours and bolometric
corrections are interpolated linearly.

For a series of absorption indices, the empirical functions of Worthey
et al. (\cite{worthey}) are used. Worthey et al. supply fitting
functions that give index strength as a function of $T_{eff}$, $\log
g$, and $[Fe/H]$.  These were obtained from observations of 460 stars,
covering a large range in the above parameters.

\subsection{Model parameters}
Once the input physics database is defined, the only free parameters
in an evolutionary synthesis calculation for an SSP are those
describing the initial mass function IMF.

The IMF can be expressed as:

\begin{equation}
\phi(m)dm \propto m^{-(1+x)}dm.
\end{equation}

where the precise exponent $x$ is observationally still somewhat
controversal. For globular clusters, Chabrier and M\'era
(\cite{chabrier_mera}) find slopes between $0.5$ and $1.5$,
independent of metallicity.  In this paper we use the standard
Salpeter IMF with a slope of $1.35$.  Variations in the slope within
the mentioned range have only a small effect on the colours and
indices.  The low mass cut-off corresponds to the hydrogen-burning
limit, which is dependent on the metallicity as shown in Chabrier and
Baraffe (\cite{chabrier}). It ranges from $0.083 M_{\odot}$ for $[M/H]
= -2.0$ to $0.075 M_{\odot}$ for $[M/H] = 0$.

\subsection{Numerical method}
\label{sec_nummethod}
The SSP models presented here are single metallicity single burst
models where SF occurs in one timestep, i.e. during the first $10^7$
yr. The exact duration of this burst does not affect the properties of
our SSPs at ages of a few Gyr.

Theoretical stellar tracks are supplied for discrete stellar masses
only.  In real stellar systems, the mass distribution is expected to
be continuous.  Using only the discrete mass grid of the track
libraries would result in severe discontinuities, since all stars of a
given mass would move to the red giant branch and die at the same
time. We would see many ''bumps'' in the luminosity and colour
evolution of the stellar population. This effect is large for SSPs
where all stars have about the same age.  For continuous star
formation rates (as in e.g. late spirals) the evolution is much
smoother as expected.

To avoid the discontinuity problem we use a \emph{Monte-Carlo}-method
to calculate the distribution of stars in the HRD at each timestep of
our evolutionary synthesis model. This method was developed by
Loxen (\cite{loxen_elba}, \cite{loxen}). For this method, no
iso\-chrones with interpolated stellar evolutionary tracks are needed.
Instead, at each timestep a large grid in stellar masses and ages is
created.  The grid is created randomly, hence the
\emph{Monte-Carlo} designation.  Each cell in this grid has a size $\Delta
m \cdot \Delta t$ in the 2-dimensional mass - time space.  The grid
ranges in mass from the lower mass to upper mass cutoff. In the case
of an SSP the grid ranges in time from zero to the end of the burst,
while for a continous SFR this would be from zero to the model age.
Each cell represents a pseudo star, which is weighted with the value
of the initial mass function (IMF) and the value of the star formation
rate (SFR) at the position of the cell. This weight is given by
\begin{equation}
\label{jolos_pet}
w = \Delta m \cdot \phi(m) \cdot \Delta t \cdot \psi(t),
\end{equation}
where $\phi(m)$ is the value of the IMF at the mass of the pseudo star
and $\psi(t)$ is the star formation rate at the time the pseudo star
is born. In the case of an SSP, the function $\psi(t)$ is constant
over the time of the burst and then equal to zero. The unit of the
value $w$ is \emph{number of stars}, although it can be
fractional. For each of these cells, the lifetime of the pseudo star
is determined by interpolation between the nearest two stellar tracks
supplied in the library. Then the timesteps of the tracks are
stretched according to this new lifetime. The cell is then split in
two parts with relative weights given by
\begin{equation}
w_1 = \frac{log(m_2) - log(m)}{log(m_2) - log(m_1)} \cdot w;\\
w_2 = w - w_1,
\end{equation}
where $m$ is the mass value of the cell, $m_1$ is the mass of the
track with the next lower mass and $m_2$ is the mass of the track with the
next higher mass.

The \emph{only} interpolation that is done is to determine the life
times and hence the duration of the individual states. \emph{No}
interpolation is done for the luminosities or the effective
temperatures which would require a precise definition of equivalent
evolutionary stages. In this way, no artificial tracks are
created. Especially in mass ranges were there is a strong dependance
of the stellar evolution on the mass this is important.

The cells in the grid \emph{do not} represent individual stars since
their weights are not necessarily unity. The weights $w$ are added in
a book-keeping list for the individual states in the tracks of the
input library to get weights $W_{i,j}$. Thereafter, the luminosities
of all states on all tracks are summed up, each weighted with the
calculated weights:
\begin{equation}
L_{total} = \sum_{i,j} W_{i,j} \cdot L_{i,j},
\end{equation}
where $L_{total}$ is the luminosity in some band, $W_{i,j}$ are the
assigned weights and $L_{i,j}$ are the luminosities of a theoretical
star from the library at $i$th mass and $j$th state.

Due to the randomly spaced grid, there is a \emph{noise} on the
results, but the finer the grid (i.e. the more cells in the grid), the
better is the signal-to-noise ratio.  For the models in this paper, we
use $200\ts 000$ masses distributed evenly in $\log$ mass and 1000
ages distributed evenly, which make up a grid of $2 \cdot 10^8$ cells.
Any further increase of the size of the grid does not give
significantly better results. For more continuous SFRs, the grid can
be coarser. Trivially the grid extends in time such that all stars are
born in the burst interval, i.e. in the first $10^7$ years. This
implies that each $\Delta t$ from equation \ref{jolos_pet} is equal to
around $10^4$ years.

For the same input physics data set, the Monte Carlo method has been
tested in detail against the standard synthesis model for SSPs (Fritze
- v.Alvensleben, \cite{fritze2}). Results for solar metallicity are
compared to those of various evolutionary synthesis codes including
the isochrone synthesis code of Bruzual \& Charlot in section
\ref{sec_compare}.

On our Linux Pentium II 300 MHz work stations a single run with a
timestep of one Gyr needs a few minutes of cpu time for the evolution
of an SSP over a Hubble time.

This method has many advantages. Models at \emph{any} time can be
calculated without the need of any interpolation between stellar
evolutionary tracks avoiding the problem of defining equivalent
evolutionary stages, ie. only the time intervals of the individual
tracks are interpolated (but not the temperatures or gravities and
therefore the colours).  It is also possible to use an arbitrary star
formation history. In the future, this code will be extended to be
able to work with arbitrary metallicities not covered by the stellar
tracks available.  Since this is an \emph{evolutionary} synthesis code
rather than a
\emph{population} synthesis code, it can also be extended to calculate
the chemical enrichment of galaxies.

\section{Variation of input parameters}

\subsection{Stellar mass loss}

The mass loss of stars, especially for low mass stars, is still not
understood very well.  A formulae often used to describe mass loss is
Reimers' law:
\begin{equation}
        \dot{M} = 1.27 \cdot 10^{-5} \cdot \eta M^{-1} L^{1.5} T_{eff}^{-2},
\end{equation}
where the parameter $\eta$ is not known \emph{a priori} and is probably a
function of metallicity.

On the horizontal branch, stars tend to be hotter the lower their mass, thus 
a \emph{larger} value of $\eta$ makes the HB population \emph{bluer}.

In this work we adopt for $\eta$ a value of $0.35$ independent of
metallicity for lack of better knowledge. For a careful study of the
effects of changing $\eta$ e.g. on the HB morphology see Buzzoni et
al. (\cite{buzzoni}). Covino \etal (\cite{covino}) present an
extensive study of the effect of different HB morphologies of
theoretical isochrones on the integrated colours.

We found that there is no signifcant effect of the HB morphology of
observed Galactic globular clusters on the optical colours.

\subsection{Non-solar abundance ratios}

There are many observations showing that the $\alpha$ - element-to-iron
ratio is larger than solar for stars with low metallicities in the
Milky Way (eg. Furmann et al. \cite{fuhrmann}). Nevertheless, most
stellar evolutionary track calculations for sub-solar metallicities
use solar abundance ratios. Salaris \etal (\cite{salaris}) show that
the effect of different $\alpha$-to-Fe element ratios on the stellar
evolution can be accounted for by appropriately changing the
conversion of global metallicity $Z$ to $[Fe/H]$.  Despite the
observed $\alpha$-enhancement for metal poor stars in the MW,
\emph{no} correction is applied since we checked that
a correction did not give better fits to the observations. In this
paper, $\log(\frac{Z}{Z_\odot}) (=: [M/H])$ was assumed to be equal
to $[Fe/H]$.

For absorption indices, any correction for $\alpha$ - enhancement
would be inappropriate for our models since we are using the index
calibrations from Worthey, which depend \emph{only} on the metallicity
and not explicitily on $[\alpha/Fe]$ - element ratios. Since Worthey's
index calibrations are determined from Galactic stars, the dependence
of $\alpha$ - enhancement on metallicity is so implicitly included.

Somewhat surprisingly we found no previous study that has examined the
effect of $[\alpha/Fe]$ on the colours at fixed temperature.  An
analysis by Kurth (\cite{kurth_mgfe}) shows that at least for solar type
stars, the colour is \emph{only} dependent on $[Fe/H]$ and
\emph{not} on $[Mg/Fe]$.

There may still be an effect of non-solar abundance ratios on the
morphology of the horizontal branch, but since our models agree with
the observations no further investigation in this direction is
attempted at the present stage.

\section{Results}
\subsection{Colours}

%

\begin{figure}
\includegraphics[width=\columnwidth]{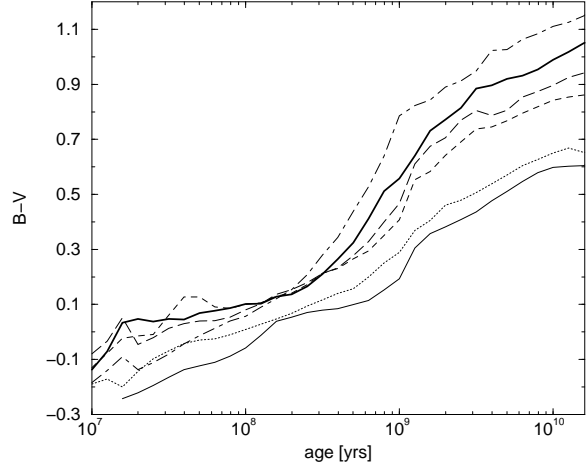}
\caption{B-V versus time with logarithmic scaling for time (solid line: Z = 0.0001, dotted line: Z = 0.0004, short-dashed line: Z = 0.004, long dashed line: Z = 0.008, thick line: Z = 0.02, dot-dashed line: Z = 0.05)}
\label{fig_logtime_bv}
\end{figure}

\begin{figure}
\includegraphics[width=\columnwidth]{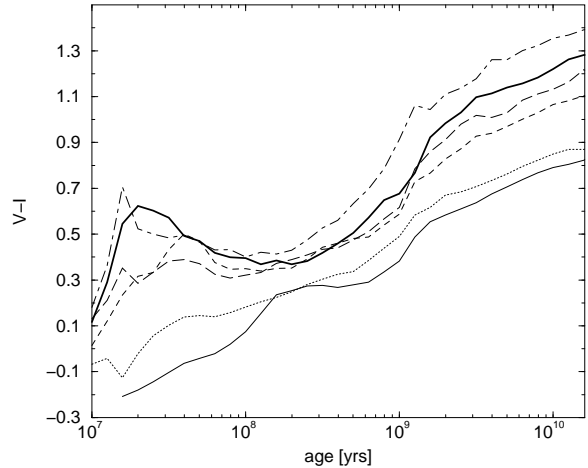}
\caption{V-I versus time with logarithmic scaling for time.}
\label{fig_logtime_vi}
\end{figure}

Figures \ref{fig_logtime_bv} and \ref{fig_logtime_vi} show the
evolution of $B-V$ and $V-I$ versus time of our SSPs on a logaritmic
time scale for all metallicities. After a few Gyrs the changes in both
colours become very slow, and the metallicity dependence becomes more
important.  Tables \ref{tab_clr1} through \ref{tab_clr3} give the time
evolution of all our broad band colours from $U$ to $K$ for 6
metallicities from $Z=0.0001$ to $Z=0.05$ for a wide range of ages.

\begin{figure}
\includegraphics[width=\columnwidth]{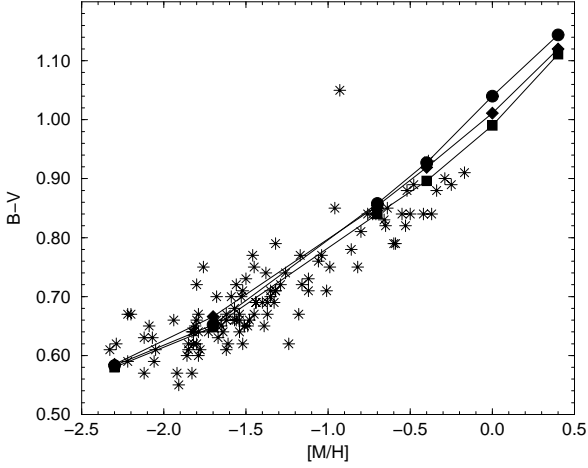}
\caption{B-V colour versus metallicity for observerd clusters \emph{(stars)} from Harris with 
$E(B-V) < 0.4$ and models at 10 \emph{(squares)},12 \emph{(diamonds)}
and 15 \emph{(circles)} Gyrs with $\eta = 0.35$. The observed colours
have been deredened.}
\label{fig_feh_bv}
\end{figure}

\begin{figure}
\includegraphics[width=\columnwidth]{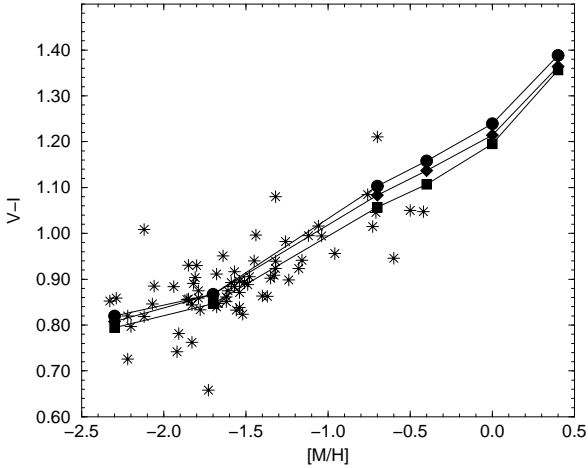}
\caption{same as figure \ref{fig_feh_bv}, but for V-I}
\label{fig_feh_vi}
\end{figure}

In Figures \ref{fig_feh_bv} and \ref{fig_feh_vi} the colours $B-V$ and
$V-I$ are shown as a function of metallicity for the models at
evolutionary ages of 10, 12 and 15 Gyr together with colors of
dereddened GCs from the McMaster catalogue (Harris \cite{harris}). In
general, we see a good agreement between the models and the observed
clusters. The large spread in colour in the observed clusters probably
arises from observational errors.

It can also be seen in both models and observations, that for very low
metallicities the colour-metallicity relation \emph{cannot} be
expressed by a simple linear function. For $[M/H] \lta -1.7$, the
relation becomes significantly flatter. The flattening is particularly
pronounced in $V-I$. The models also show that the $(V-I)$-metallicity
relation steepens for $[M/H] > 0$.  Very often the metallicity is
calculated for globular cluster systems in other galaxies from $V-I$
using a simple linear regression line calculated from Galactic
GCs. The models show that this approach is dangerous for metallicities
higher than those of Galactic GCs. A quadratic or higher order fit
would not improve things if calculated for low metallicities and
applied to higher metallicities as it would include further
uncertainties. Therefore a theoretical calibration like the one
provided here is to be preferred.

\subsection{Indices}
\begin{figure}
\includegraphics[width=\columnwidth]{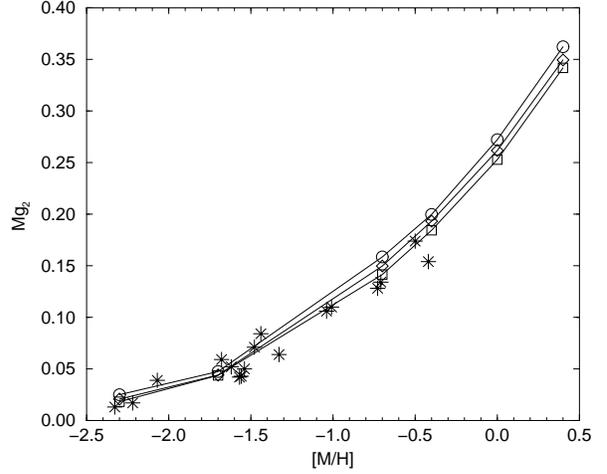}
\caption{Mg$_2$ versus metallicity for observerd clusters \emph{(stars)} and models at 10 \emph{(squares)},12 \emph{(diamonds)} and 15 \emph{(circles)} Gyrs with $\eta = 0.35$. The observed indices are from Burstein, their metallicities from Harris.}
\label{fig_feh_mg2}
\end{figure}

\begin{figure}
\includegraphics[width=\columnwidth]{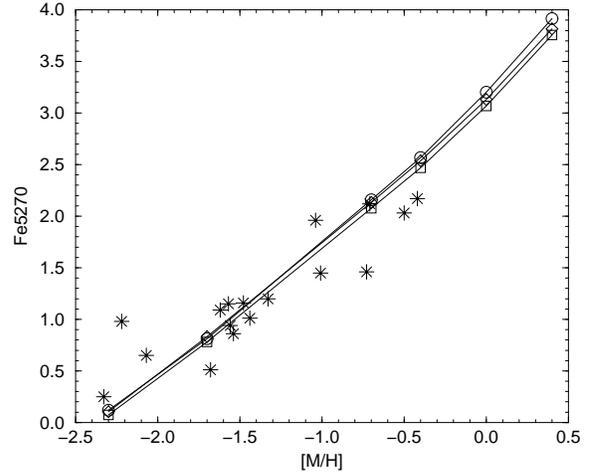}
\caption{same as figure \ref{fig_feh_mg2}, but for Fe5270}
\label{fig_feh_fe5270}
\end{figure}

\begin{figure}
\includegraphics[width=\columnwidth]{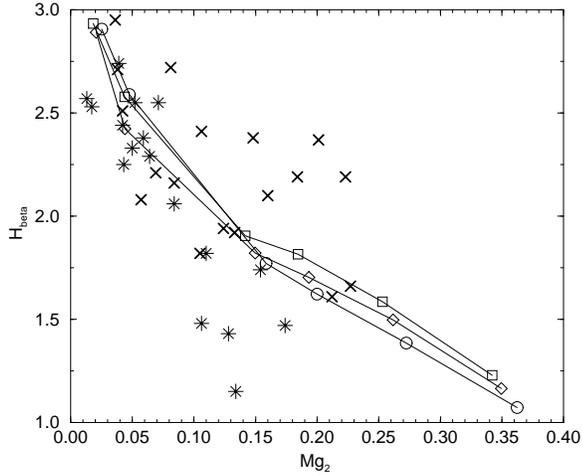}
\caption{The $H_\beta$ index against Mg$_2$ index for observations of Galactic GCs (stars), M31 GCs (crosses) and models.}
\label{fig_mg2_hbeta}
\end{figure}

\begin{figure}
\includegraphics[width=\columnwidth]{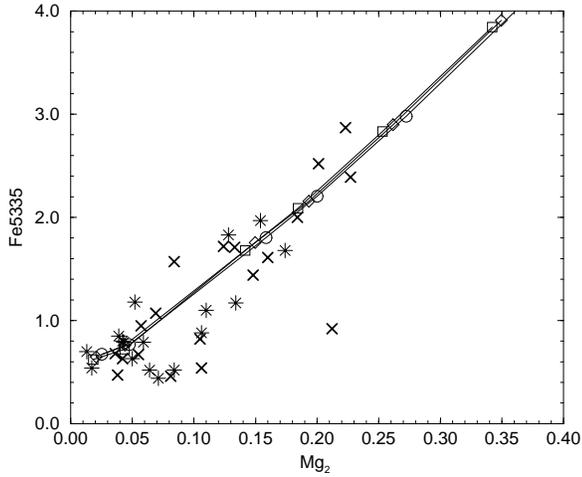}
\caption{Same as figure \ref{fig_mg2_hbeta} but for the Fe5335 index.}
\label{fig_mg2_fe5335}
\end{figure}

\begin{figure}
\includegraphics[width=\columnwidth]{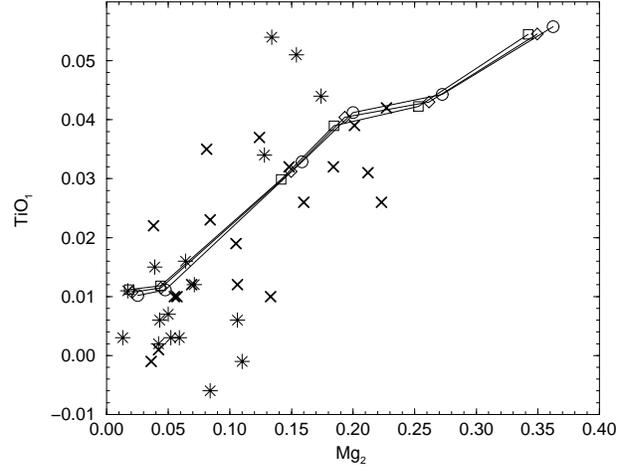}
\caption{Same as figure \ref{fig_mg2_hbeta} but for the TiO$_1$ index.}
\label{fig_mg2_tio1}
\end{figure}

\begin{figure}
\includegraphics[width=\columnwidth]{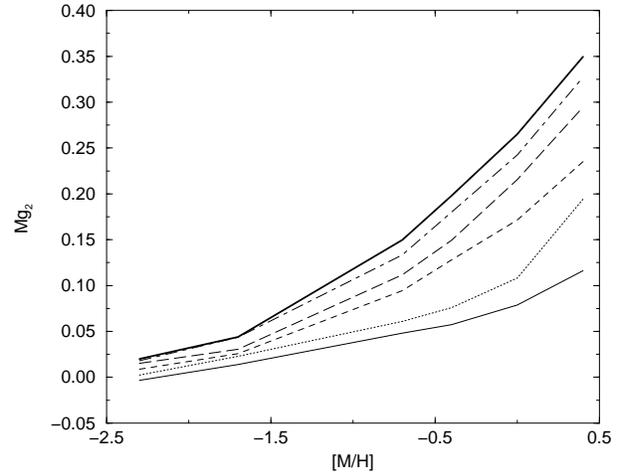}
\caption{The $Mg_2$ index against metallicity for various ages. Thin solid line: 0.5 Gyrs, dotted line: 1 Gyr, dashed: 2 Gyrs, long dashed: 4 Gyrs, dot dashed: 8 Gyrs, thick solid line: 12 Gyrs.}
\label{fig_met_mg2_ages}
\end{figure}

In Figures \ref{fig_feh_mg2} and \ref{fig_feh_fe5270} we plot against
metallicity the indices Mg$_2$ and Fe5270 from our models and those of
Galactic clusters from Burstein (\cite{burstein}).
The agreement of the models with the observations is very good over
the metalicity range covered by the data.  Model calibrations of both
Mg$_2$ and Fe5270 as functions of metallicity are almost independent
of age for ages close to a Hubble time.

Figure \ref{fig_mg2_hbeta} shows the $H_\beta$ index against the
Mg$_2$ for Galactic GCs and clusters from M31. As can be seen in the
figure the $H_\beta$ index is higher for higher metallicity for the M
31 GCs than for the Milky Way clusters.  This was already noted by
Burstein (\cite{burstein}) though the reason for this discrepancy is
still unknown.  The models fall right between the two groups.  It can
also be seen that the H$_\beta$ to Mg$_2$ relation is dependent
largely on age.  If it were not for the large spread present in the
H$_\beta$ observations this index would be a good tool to disentangle
age from metallicity.

The Fe5335 and TiO$_1$ are plotted against Mg$_2$ for observations
from Burstein in Figures \ref{fig_mg2_fe5335} and \ref{fig_mg2_tio1}
respectively.  Both the relations between the Fe5335 and TiO$_1$
indices and Mg$_2$ are virtually independent of model age for ages
close to a Hubble time.

In Figure \ref{fig_met_mg2_ages} we show the model index Mg$_2$
against metallicity for a wide range of ages from $0.5$ to $12$
Gyrs.

\subsection{Comparison with other authors}
\label{sec_compare}
\begin{figure}
\includegraphics[width=\columnwidth]{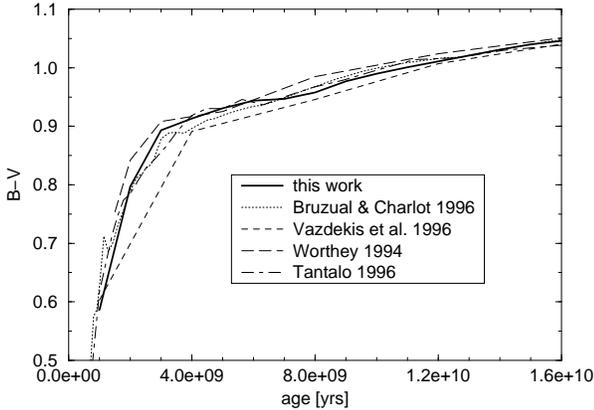}
\caption{Comparison of $B-V$ against age predicted with our models with those of other authors}
\label{fig_bv_others}
\end{figure}

\begin{figure}
\includegraphics[width=\columnwidth]{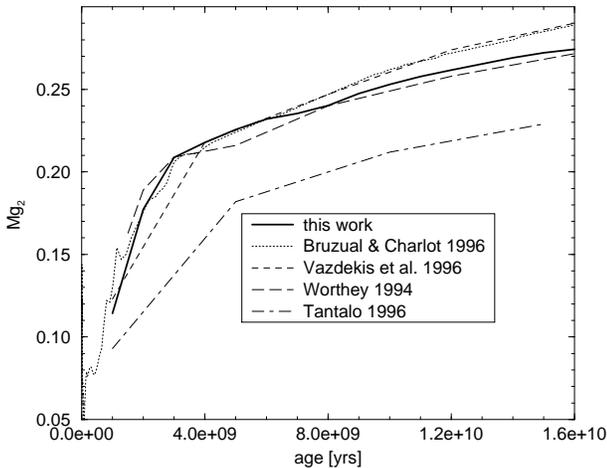}
\caption{Comparison of the Mg$_2$ index against age as predicted with our models to those of other authors}
\label{fig_mg2_others}
\end{figure}

We compared our results with those of Bruzual \& Charlot (BC96)
(\cite{bruzual}), Vazdekis \etal (\cite{vazdekis}), Worthey
(\cite{worthey1}), Tantalo \etal (\cite{tantalo}) (for $B-V$) and
Tantalo \etal (\cite{tantalo2}) (for Mg$_2$). With the exception of
the models from Worthey, all models are based on the Padua tracks.

Our $B-V$ - colours are very close to those of the BC96 models. This
is surprising since they also calibrated their colours with the
library from Lejeune.  The models from Vazdekis \etal are bluer at all
times. They use a different empirical calibration.  The $B-V$ -
colours for all models seem to converge at high ages.

The calibration of the indices are from Worthey \cite{worthey} for all
models with the exception of the models from Tantalo \etal
(\cite{tantalo2}), who use the empirical calibrations from Borges \etal
(\cite{borges}) for Mg$_2$ which depend on the [Mg/Fe] ratio.

Our Mg$_2$ indices are lower than those of BC96 and Vazdekis at high
ages by less than 0.02 mag, but more close to those of
Worthey. Probably due to the different calibrations for their indices,
the Mg$_2$ indices of Tantalo \etal are lower than those of all the
other models considered here.

\section{Summary}
In this work we present Monte-Carlo evolutionary synthesis models for
SSPs which cover a wide range in metallicity, from $Z=0.0001$ to $Z=0.05$, using
most recent and complete sets of input physics: stellar
evolutionary tracks for stellar masses from $0.08 M_\odot$ to $120
M_\odot$, including post - helium flash evolution and mass loss, model
atmosphere libraries also covering late stellar types and giving
colours from $U$ to $K$ in agreement with observations and empirical
calibrations for a series of absorption indices.

We obtain theoretical calibrations of colours and indices in terms of
metallicity which for model ages of 10-15 Gys agree closely with
observations of GCs. The theoretical calibrations extend beyond the
range of observed GCs, i.e. to a metallicity up to $[M/H] \leq 0.4$.
Moreover, our models provide these theoretical calibrations for
all ages from cluster formation to 15 Gyrs and thus can also be
applied in the interpretation of young star cluster systems observed
in many interacting and starburst galaxies.

The complete model files are available via WWW on {\tt
http://www.uni-sw.gwdg.de/\~\ okurth/ssp.html}. There are also models
with different parameters for the IMF and for mass loss available.

\begin{acknowledgements}
We thank Thibault Lejeune for providing us his library of theoretical
stellar spectra, Isabelle Baraffe for providing us with her tracks for
the low mass stars and our anonymous referee for valuable and
instructive suggestions. This work was supported by the German
\emph{Deut\-sche For\-schungs\-ge\-mein\-schaft, DFG\/} project
number FR~916/3-1.
\end{acknowledgements}

\begin{table}
\caption{The model colours. Time is in years. U, B and V are in the Johnson system, R and I in the Cousins system, K as in Bessell and Brett \cite{bessell}.}
\label{tab_clr1}
\begin{tabular}{r|r|r|r|r|r}
\hline
time& $U-B$ & $B-V$ & $V-R$ & $V-I$ & $V-K$ \\
\hline
\multicolumn{6}{c}{Z = 0.0001}\\
\hline
$1.0 \cdot 10^{7}$ & $-0.80$ & $0.05$ & $0.14$ & $0.34$ & $0.99$ \\
$2.0 \cdot 10^{7}$ & $-0.90$ & $-0.22$ & $-0.09$ & $-0.18$ & $-0.41$ \\
$3.0 \cdot 10^{7}$ & $-0.76$ & $-0.17$ & $-0.06$ & $-0.11$ & $-0.19$ \\
$4.0 \cdot 10^{7}$ & $-0.65$ & $-0.14$ & $-0.04$ & $-0.06$ & $-0.03$ \\
$5.0 \cdot 10^{7}$ & $-0.60$ & $-0.12$ & $-0.03$ & $-0.04$ & $0.03$ \\
$6.0 \cdot 10^{7}$ & $-0.57$ & $-0.11$ & $-0.03$ & $-0.03$ & $0.06$ \\
$7.0 \cdot 10^{7}$ & $-0.55$ & $-0.10$ & $-0.02$ & $-0.01$ & $0.15$ \\
$8.0 \cdot 10^{7}$ & $-0.53$ & $-0.09$ & $-0.01$ & $0.02$ & $0.23$ \\
$9.0 \cdot 10^{7}$ & $-0.51$ & $-0.07$ & $0.01$ & $0.05$ & $0.32$ \\
$1.0 \cdot 10^{8}$ & $-0.49$ & $-0.06$ & $0.02$ & $0.07$ & $0.40$ \\
$2.0 \cdot 10^{8}$ & $-0.39$ & $0.06$ & $0.10$ & $0.25$ & $0.88$ \\
$3.0 \cdot 10^{8}$ & $-0.30$ & $0.08$ & $0.11$ & $0.27$ & $0.91$ \\
$4.0 \cdot 10^{8}$ & $-0.21$ & $0.08$ & $0.11$ & $0.27$ & $0.89$ \\
$5.0 \cdot 10^{8}$ & $-0.17$ & $0.10$ & $0.11$ & $0.28$ & $0.91$ \\
$6.0 \cdot 10^{8}$ & $-0.12$ & $0.11$ & $0.12$ & $0.29$ & $0.91$ \\
$7.0 \cdot 10^{8}$ & $-0.09$ & $0.11$ & $0.12$ & $0.28$ & $0.88$ \\
$8.0 \cdot 10^{8}$ & $-0.06$ & $0.15$ & $0.14$ & $0.34$ & $0.99$ \\
$9.0 \cdot 10^{8}$ & $-0.02$ & $0.10$ & $0.09$ & $0.23$ & $0.74$ \\
$1.0 \cdot 10^{9}$ & $-0.08$ & $0.19$ & $0.16$ & $0.38$ & $1.11$ \\
$1.0 \cdot 10^{9}$ & $-0.08$ & $0.19$ & $0.16$ & $0.38$ & $1.11$ \\
$2.0 \cdot 10^{9}$ & $-0.02$ & $0.38$ & $0.27$ & $0.58$ & $1.51$ \\
$3.0 \cdot 10^{9}$ & $-0.01$ & $0.43$ & $0.29$ & $0.63$ & $1.59$ \\
$4.0 \cdot 10^{9}$ & $-0.01$ & $0.48$ & $0.32$ & $0.68$ & $1.67$ \\
$5.0 \cdot 10^{9}$ & $-0.01$ & $0.51$ & $0.34$ & $0.71$ & $1.73$ \\
$6.0 \cdot 10^{9}$ & $-0.02$ & $0.54$ & $0.35$ & $0.73$ & $1.77$ \\
$7.0 \cdot 10^{9}$ & $-0.02$ & $0.56$ & $0.36$ & $0.75$ & $1.81$ \\
$8.0 \cdot 10^{9}$ & $-0.03$ & $0.58$ & $0.37$ & $0.77$ & $1.84$ \\
$9.0 \cdot 10^{9}$ & $-0.03$ & $0.59$ & $0.38$ & $0.78$ & $1.86$ \\
$1.0 \cdot 10^{10}$ & $-0.02$ & $0.60$ & $0.38$ & $0.79$ & $1.88$ \\
$1.1 \cdot 10^{10}$ & $-0.02$ & $0.60$ & $0.38$ & $0.80$ & $1.90$ \\
$1.2 \cdot 10^{10}$ & $-0.01$ & $0.60$ & $0.39$ & $0.80$ & $1.91$ \\
$1.3 \cdot 10^{10}$ & $0.00$ & $0.60$ & $0.39$ & $0.81$ & $1.93$ \\
$1.4 \cdot 10^{10}$ & $0.00$ & $0.60$ & $0.39$ & $0.81$ & $1.93$ \\
$1.5 \cdot 10^{10}$ & $0.00$ & $0.60$ & $0.39$ & $0.81$ & $1.95$ \\
$1.6 \cdot 10^{10}$ & $-0.01$ & $0.59$ & $0.39$ & $0.82$ & $1.95$ \\
\hline
\end{tabular}
\end{table}

\begin{table}
\caption{continuation of Table \ref{tab_clr1} for Z = 0.0004.}
\label{tab_clr2}
\begin{tabular}{r|r|r|r|r|r}
\hline
time& $U-B$ & $B-V$ & $V-R$ & $V-I$ & $V-K$ \\
\hline
\multicolumn{6}{c}{Z = 0.0004}\\
\hline
$1.0 \cdot 10^{7}$ & $-0.96$ & $-0.19$ & $-0.04$ & $-0.07$ & $-0.01$ \\
$2.0 \cdot 10^{7}$ & $-0.80$ & $-0.14$ & $-0.03$ & $-0.02$ & $0.21$ \\
$3.0 \cdot 10^{7}$ & $-0.67$ & $-0.08$ & $0.02$ & $0.09$ & $0.55$ \\
$4.0 \cdot 10^{7}$ & $-0.59$ & $-0.04$ & $0.05$ & $0.14$ & $0.66$ \\
$5.0 \cdot 10^{7}$ & $-0.54$ & $-0.03$ & $0.05$ & $0.14$ & $0.66$ \\
$6.0 \cdot 10^{7}$ & $-0.51$ & $-0.03$ & $0.05$ & $0.14$ & $0.62$ \\
$7.0 \cdot 10^{7}$ & $-0.47$ & $-0.02$ & $0.05$ & $0.15$ & $0.64$ \\
$8.0 \cdot 10^{7}$ & $-0.44$ & $-0.01$ & $0.06$ & $0.16$ & $0.66$ \\
$9.0 \cdot 10^{7}$ & $-0.41$ & $0.00$ & $0.06$ & $0.17$ & $0.69$ \\
$1.0 \cdot 10^{8}$ & $-0.39$ & $0.01$ & $0.07$ & $0.18$ & $0.71$ \\
$2.0 \cdot 10^{8}$ & $-0.25$ & $0.07$ & $0.10$ & $0.25$ & $0.85$ \\
$3.0 \cdot 10^{8}$ & $-0.16$ & $0.11$ & $0.12$ & $0.30$ & $0.95$ \\
$4.0 \cdot 10^{8}$ & $-0.11$ & $0.14$ & $0.14$ & $0.33$ & $0.99$ \\
$5.0 \cdot 10^{8}$ & $-0.10$ & $0.16$ & $0.14$ & $0.34$ & $1.00$ \\
$6.0 \cdot 10^{8}$ & $-0.09$ & $0.19$ & $0.16$ & $0.37$ & $1.06$ \\
$7.0 \cdot 10^{8}$ & $-0.09$ & $0.22$ & $0.18$ & $0.41$ & $1.12$ \\
$8.0 \cdot 10^{8}$ & $-0.09$ & $0.25$ & $0.20$ & $0.44$ & $1.17$ \\
$9.0 \cdot 10^{8}$ & $-0.09$ & $0.27$ & $0.21$ & $0.47$ & $1.23$ \\
$1.0 \cdot 10^{9}$ & $-0.10$ & $0.29$ & $0.22$ & $0.49$ & $1.26$ \\
$1.0 \cdot 10^{9}$ & $-0.10$ & $0.29$ & $0.22$ & $0.49$ & $1.26$ \\
$2.0 \cdot 10^{9}$ & $0.01$ & $0.46$ & $0.32$ & $0.67$ & $1.69$ \\
$3.0 \cdot 10^{9}$ & $0.01$ & $0.50$ & $0.33$ & $0.70$ & $1.74$ \\
$4.0 \cdot 10^{9}$ & $0.00$ & $0.54$ & $0.35$ & $0.73$ & $1.80$ \\
$5.0 \cdot 10^{9}$ & $-0.01$ & $0.57$ & $0.36$ & $0.76$ & $1.84$ \\
$6.0 \cdot 10^{9}$ & $-0.01$ & $0.60$ & $0.38$ & $0.79$ & $1.89$ \\
$7.0 \cdot 10^{9}$ & $-0.02$ & $0.61$ & $0.39$ & $0.81$ & $1.92$ \\
$8.0 \cdot 10^{9}$ & $-0.02$ & $0.63$ & $0.40$ & $0.82$ & $1.95$ \\
$9.0 \cdot 10^{9}$ & $-0.02$ & $0.64$ & $0.41$ & $0.84$ & $1.98$ \\
$1.0 \cdot 10^{10}$ & $-0.02$ & $0.65$ & $0.41$ & $0.85$ & $2.00$ \\
$1.1 \cdot 10^{10}$ & $-0.02$ & $0.66$ & $0.42$ & $0.85$ & $2.01$ \\
$1.2 \cdot 10^{10}$ & $-0.01$ & $0.67$ & $0.42$ & $0.87$ & $2.04$ \\
$1.3 \cdot 10^{10}$ & $-0.01$ & $0.67$ & $0.42$ & $0.87$ & $2.05$ \\
$1.4 \cdot 10^{10}$ & $0.00$ & $0.67$ & $0.43$ & $0.88$ & $2.06$ \\
$1.5 \cdot 10^{10}$ & $0.01$ & $0.65$ & $0.42$ & $0.87$ & $2.05$ \\
$1.6 \cdot 10^{10}$ & $0.02$ & $0.66$ & $0.42$ & $0.88$ & $2.07$ \\
\hline
\end{tabular}
\end{table}

\begin{table}
\caption{continuation of Table \ref{tab_clr1} for Z = 0.004.}
\label{tab_clr3}
\begin{tabular}{r|r|r|r|r|r}
\hline
time& $U-B$ & $B-V$ & $V-R$ & $V-I$ & $V-K$ \\
\hline
\multicolumn{6}{c}{Z = 0.004}\\
\hline
$1.0 \cdot 10^{7}$ & $-0.85$ & $-0.13$ & $-0.01$ & $0.01$ & $0.32$ \\
$2.0 \cdot 10^{7}$ & $-0.72$ & $-0.01$ & $0.12$ & $0.32$ & $1.26$ \\
$3.0 \cdot 10^{7}$ & $-0.61$ & $0.04$ & $0.16$ & $0.41$ & $1.49$ \\
$4.0 \cdot 10^{7}$ & $-0.50$ & $0.13$ & $0.21$ & $0.50$ & $1.64$ \\
$5.0 \cdot 10^{7}$ & $-0.44$ & $0.13$ & $0.19$ & $0.47$ & $1.56$ \\
$6.0 \cdot 10^{7}$ & $-0.42$ & $0.10$ & $0.16$ & $0.39$ & $1.38$ \\
$7.0 \cdot 10^{7}$ & $-0.40$ & $0.09$ & $0.15$ & $0.36$ & $1.31$ \\
$8.0 \cdot 10^{7}$ & $-0.38$ & $0.09$ & $0.14$ & $0.35$ & $1.26$ \\
$9.0 \cdot 10^{7}$ & $-0.36$ & $0.09$ & $0.14$ & $0.35$ & $1.26$ \\
$1.0 \cdot 10^{8}$ & $-0.33$ & $0.10$ & $0.14$ & $0.35$ & $1.26$ \\
$2.0 \cdot 10^{8}$ & $-0.21$ & $0.14$ & $0.15$ & $0.35$ & $1.20$ \\
$3.0 \cdot 10^{8}$ & $-0.13$ & $0.20$ & $0.18$ & $0.43$ & $1.32$ \\
$4.0 \cdot 10^{8}$ & $-0.08$ & $0.23$ & $0.20$ & $0.46$ & $1.37$ \\
$5.0 \cdot 10^{8}$ & $-0.02$ & $0.27$ & $0.21$ & $0.48$ & $1.39$ \\
$6.0 \cdot 10^{8}$ & $0.02$ & $0.28$ & $0.22$ & $0.48$ & $1.37$ \\
$7.0 \cdot 10^{8}$ & $0.06$ & $0.32$ & $0.23$ & $0.51$ & $1.42$ \\
$8.0 \cdot 10^{8}$ & $0.09$ & $0.35$ & $0.25$ & $0.54$ & $1.48$ \\
$9.0 \cdot 10^{8}$ & $0.11$ & $0.39$ & $0.27$ & $0.57$ & $1.54$ \\
$1.0 \cdot 10^{9}$ & $0.10$ & $0.41$ & $0.27$ & $0.59$ & $1.54$ \\
$1.0 \cdot 10^{9}$ & $0.10$ & $0.41$ & $0.27$ & $0.59$ & $1.54$ \\
$2.0 \cdot 10^{9}$ & $0.15$ & $0.64$ & $0.40$ & $0.82$ & $2.08$ \\
$3.0 \cdot 10^{9}$ & $0.18$ & $0.74$ & $0.45$ & $0.92$ & $2.27$ \\
$4.0 \cdot 10^{9}$ & $0.16$ & $0.75$ & $0.46$ & $0.94$ & $2.30$ \\
$5.0 \cdot 10^{9}$ & $0.17$ & $0.77$ & $0.47$ & $0.97$ & $2.36$ \\
$6.0 \cdot 10^{9}$ & $0.18$ & $0.79$ & $0.48$ & $1.00$ & $2.42$ \\
$7.0 \cdot 10^{9}$ & $0.20$ & $0.81$ & $0.50$ & $1.02$ & $2.46$ \\
$8.0 \cdot 10^{9}$ & $0.21$ & $0.82$ & $0.50$ & $1.03$ & $2.48$ \\
$9.0 \cdot 10^{9}$ & $0.22$ & $0.83$ & $0.51$ & $1.05$ & $2.52$ \\
$1.0 \cdot 10^{10}$ & $0.23$ & $0.84$ & $0.51$ & $1.06$ & $2.54$ \\
$1.1 \cdot 10^{10}$ & $0.24$ & $0.85$ & $0.52$ & $1.07$ & $2.56$ \\
$1.2 \cdot 10^{10}$ & $0.24$ & $0.85$ & $0.53$ & $1.08$ & $2.59$ \\
$1.3 \cdot 10^{10}$ & $0.26$ & $0.86$ & $0.53$ & $1.10$ & $2.64$ \\
$1.4 \cdot 10^{10}$ & $0.25$ & $0.86$ & $0.53$ & $1.10$ & $2.63$ \\
$1.5 \cdot 10^{10}$ & $0.25$ & $0.86$ & $0.53$ & $1.10$ & $2.64$ \\
$1.6 \cdot 10^{10}$ & $0.26$ & $0.86$ & $0.54$ & $1.11$ & $2.66$ \\
\hline
\end{tabular}
\end{table}

\begin{table}
\caption{continuation of Table \ref{tab_clr1} for Z = 0.008.}
\label{tab_clr4}
\begin{tabular}{r|r|r|r|r|r}
\hline
time& $U-B$ & $B-V$ & $V-R$ & $V-I$ & $V-K$ \\
\hline
\multicolumn{6}{c}{Z = 0.008}\\
\hline
$1.0 \cdot 10^{7}$ & $-0.88$ & $-0.08$ & $0.04$ & $0.13$ & $0.66$ \\
$2.0 \cdot 10^{7}$ & $-0.71$ & $-0.04$ & $0.10$ & $0.28$ & $1.31$ \\
$3.0 \cdot 10^{7}$ & $-0.59$ & $0.01$ & $0.14$ & $0.38$ & $1.51$ \\
$4.0 \cdot 10^{7}$ & $-0.51$ & $0.03$ & $0.14$ & $0.39$ & $1.52$ \\
$5.0 \cdot 10^{7}$ & $-0.46$ & $0.04$ & $0.14$ & $0.37$ & $1.45$ \\
$6.0 \cdot 10^{7}$ & $-0.43$ & $0.04$ & $0.13$ & $0.33$ & $1.32$ \\
$7.0 \cdot 10^{7}$ & $-0.39$ & $0.05$ & $0.12$ & $0.32$ & $1.27$ \\
$8.0 \cdot 10^{7}$ & $-0.36$ & $0.05$ & $0.12$ & $0.31$ & $1.22$ \\
$9.0 \cdot 10^{7}$ & $-0.34$ & $0.07$ & $0.12$ & $0.31$ & $1.22$ \\
$1.0 \cdot 10^{8}$ & $-0.32$ & $0.08$ & $0.13$ & $0.32$ & $1.23$ \\
$2.0 \cdot 10^{8}$ & $-0.21$ & $0.15$ & $0.16$ & $0.39$ & $1.29$ \\
$3.0 \cdot 10^{8}$ & $-0.10$ & $0.20$ & $0.19$ & $0.43$ & $1.35$ \\
$4.0 \cdot 10^{8}$ & $-0.01$ & $0.23$ & $0.19$ & $0.44$ & $1.37$ \\
$5.0 \cdot 10^{8}$ & $0.06$ & $0.28$ & $0.21$ & $0.47$ & $1.42$ \\
$6.0 \cdot 10^{8}$ & $0.10$ & $0.32$ & $0.23$ & $0.50$ & $1.46$ \\
$7.0 \cdot 10^{8}$ & $0.13$ & $0.36$ & $0.25$ & $0.53$ & $1.50$ \\
$8.0 \cdot 10^{8}$ & $0.15$ & $0.40$ & $0.27$ & $0.57$ & $1.56$ \\
$9.0 \cdot 10^{8}$ & $0.17$ & $0.46$ & $0.29$ & $0.62$ & $1.66$ \\
$1.0 \cdot 10^{9}$ & $0.16$ & $0.47$ & $0.29$ & $0.62$ & $1.63$ \\
$1.0 \cdot 10^{9}$ & $0.16$ & $0.47$ & $0.29$ & $0.62$ & $1.63$ \\
$2.0 \cdot 10^{9}$ & $0.18$ & $0.71$ & $0.44$ & $0.91$ & $2.31$ \\
$3.0 \cdot 10^{9}$ & $0.24$ & $0.80$ & $0.49$ & $1.01$ & $2.52$ \\
$4.0 \cdot 10^{9}$ & $0.22$ & $0.79$ & $0.48$ & $1.01$ & $2.54$ \\
$5.0 \cdot 10^{9}$ & $0.23$ & $0.81$ & $0.49$ & $1.03$ & $2.60$ \\
$6.0 \cdot 10^{9}$ & $0.28$ & $0.85$ & $0.52$ & $1.08$ & $2.71$ \\
$7.0 \cdot 10^{9}$ & $0.30$ & $0.86$ & $0.52$ & $1.09$ & $2.74$ \\
$8.0 \cdot 10^{9}$ & $0.32$ & $0.88$ & $0.53$ & $1.11$ & $2.77$ \\
$9.0 \cdot 10^{9}$ & $0.33$ & $0.89$ & $0.53$ & $1.11$ & $2.76$ \\
$1.0 \cdot 10^{10}$ & $0.35$ & $0.90$ & $0.54$ & $1.13$ & $2.81$ \\
$1.1 \cdot 10^{10}$ & $0.37$ & $0.91$ & $0.55$ & $1.15$ & $2.84$ \\
$1.2 \cdot 10^{10}$ & $0.39$ & $0.92$ & $0.56$ & $1.17$ & $2.89$ \\
$1.3 \cdot 10^{10}$ & $0.41$ & $0.93$ & $0.56$ & $1.16$ & $2.87$ \\
$1.4 \cdot 10^{10}$ & $0.42$ & $0.94$ & $0.56$ & $1.19$ & $2.94$ \\
$1.5 \cdot 10^{10}$ & $0.42$ & $0.93$ & $0.56$ & $1.19$ & $2.96$ \\
$1.6 \cdot 10^{10}$ & $0.42$ & $0.94$ & $0.57$ & $1.20$ & $2.94$ \\
\hline
\end{tabular}
\end{table}

\begin{table}
\caption{continuation of Table \ref{tab_clr1} for Z = 0.02.}
\label{tab_clr5}
\begin{tabular}{r|r|r|r|r|r}
\hline
time& $U-B$ & $B-V$ & $V-R$ & $V-I$ & $V-K$ \\
\hline
\multicolumn{6}{c}{Z = 0.02}\\
\hline
$1.0 \cdot 10^{7}$ & $-0.93$ & $-0.14$ & $0.02$ & $0.12$ & $0.95$ \\
$2.0 \cdot 10^{7}$ & $-0.70$ & $0.05$ & $0.22$ & $0.62$ & $2.23$ \\
$3.0 \cdot 10^{7}$ & $-0.64$ & $0.04$ & $0.20$ & $0.58$ & $2.16$ \\
$4.0 \cdot 10^{7}$ & $-0.56$ & $0.04$ & $0.18$ & $0.49$ & $1.90$ \\
$5.0 \cdot 10^{7}$ & $-0.50$ & $0.07$ & $0.18$ & $0.47$ & $1.78$ \\
$6.0 \cdot 10^{7}$ & $-0.46$ & $0.07$ & $0.17$ & $0.42$ & $1.61$ \\
$7.0 \cdot 10^{7}$ & $-0.42$ & $0.09$ & $0.17$ & $0.42$ & $1.59$ \\
$8.0 \cdot 10^{7}$ & $-0.40$ & $0.09$ & $0.16$ & $0.40$ & $1.54$ \\
$9.0 \cdot 10^{7}$ & $-0.37$ & $0.09$ & $0.16$ & $0.40$ & $1.53$ \\
$1.0 \cdot 10^{8}$ & $-0.34$ & $0.10$ & $0.16$ & $0.40$ & $1.53$ \\
$2.0 \cdot 10^{8}$ & $-0.15$ & $0.14$ & $0.15$ & $0.37$ & $1.41$ \\
$3.0 \cdot 10^{8}$ & $-0.01$ & $0.20$ & $0.18$ & $0.41$ & $1.45$ \\
$4.0 \cdot 10^{8}$ & $0.10$ & $0.27$ & $0.20$ & $0.46$ & $1.50$ \\
$5.0 \cdot 10^{8}$ & $0.16$ & $0.32$ & $0.23$ & $0.51$ & $1.55$ \\
$6.0 \cdot 10^{8}$ & $0.20$ & $0.40$ & $0.26$ & $0.56$ & $1.63$ \\
$7.0 \cdot 10^{8}$ & $0.21$ & $0.46$ & $0.29$ & $0.60$ & $1.69$ \\
$8.0 \cdot 10^{8}$ & $0.21$ & $0.51$ & $0.31$ & $0.64$ & $1.76$ \\
$9.0 \cdot 10^{8}$ & $0.22$ & $0.56$ & $0.33$ & $0.68$ & $1.82$ \\
$1.0 \cdot 10^{9}$ & $0.18$ & $0.56$ & $0.33$ & $0.68$ & $1.80$ \\
$1.0 \cdot 10^{9}$ & $0.18$ & $0.56$ & $0.33$ & $0.68$ & $1.80$ \\
$2.0 \cdot 10^{9}$ & $0.26$ & $0.77$ & $0.47$ & $0.98$ & $2.58$ \\
$3.0 \cdot 10^{9}$ & $0.37$ & $0.87$ & $0.52$ & $1.09$ & $2.85$ \\
$4.0 \cdot 10^{9}$ & $0.40$ & $0.90$ & $0.54$ & $1.11$ & $2.93$ \\
$5.0 \cdot 10^{9}$ & $0.43$ & $0.92$ & $0.55$ & $1.14$ & $2.99$ \\
$6.0 \cdot 10^{9}$ & $0.46$ & $0.94$ & $0.56$ & $1.16$ & $3.05$ \\
$7.0 \cdot 10^{9}$ & $0.47$ & $0.94$ & $0.56$ & $1.17$ & $3.06$ \\
$8.0 \cdot 10^{9}$ & $0.50$ & $0.95$ & $0.57$ & $1.18$ & $3.08$ \\
$9.0 \cdot 10^{9}$ & $0.53$ & $0.97$ & $0.58$ & $1.21$ & $3.14$ \\
$1.0 \cdot 10^{10}$ & $0.56$ & $0.99$ & $0.58$ & $1.22$ & $3.18$ \\
$1.1 \cdot 10^{10}$ & $0.59$ & $1.00$ & $0.59$ & $1.24$ & $3.25$ \\
$1.2 \cdot 10^{10}$ & $0.61$ & $1.01$ & $0.60$ & $1.25$ & $3.23$ \\
$1.3 \cdot 10^{10}$ & $0.64$ & $1.02$ & $0.60$ & $1.25$ & $3.21$ \\
$1.4 \cdot 10^{10}$ & $0.66$ & $1.03$ & $0.61$ & $1.27$ & $3.29$ \\
$1.5 \cdot 10^{10}$ & $0.68$ & $1.04$ & $0.61$ & $1.28$ & $3.27$ \\
$1.6 \cdot 10^{10}$ & $0.70$ & $1.05$ & $0.62$ & $1.29$ & $3.34$ \\
\hline
\end{tabular}
\end{table}

\begin{table}
\caption{continuation of Table \ref{tab_clr1} for Z = 0.05.}
\label{tab_clr6}
\begin{tabular}{r|r|r|r|r|r}
\hline
time& $U-B$ & $B-V$ & $V-R$ & $V-I$ & $V-K$ \\
\hline
\multicolumn{6}{c}{Z = 0.05}\\
\hline
$1.0 \cdot 10^{7}$ & $-0.95$ & $-0.18$ & $-0.01$ & $0.18$ & $1.74$ \\
$2.0 \cdot 10^{7}$ & $-0.77$ & $-0.14$ & $0.04$ & $0.52$ & $3.07$ \\
$3.0 \cdot 10^{7}$ & $-0.65$ & $-0.09$ & $0.07$ & $0.49$ & $2.64$ \\
$4.0 \cdot 10^{7}$ & $-0.56$ & $-0.04$ & $0.11$ & $0.50$ & $2.38$ \\
$5.0 \cdot 10^{7}$ & $-0.50$ & $-0.01$ & $0.13$ & $0.46$ & $2.13$ \\
$6.0 \cdot 10^{7}$ & $-0.45$ & $0.01$ & $0.14$ & $0.44$ & $1.99$ \\
$7.0 \cdot 10^{7}$ & $-0.40$ & $0.02$ & $0.14$ & $0.43$ & $1.93$ \\
$8.0 \cdot 10^{7}$ & $-0.36$ & $0.04$ & $0.15$ & $0.43$ & $1.89$ \\
$9.0 \cdot 10^{7}$ & $-0.33$ & $0.05$ & $0.15$ & $0.41$ & $1.81$ \\
$1.0 \cdot 10^{8}$ & $-0.30$ & $0.06$ & $0.15$ & $0.40$ & $1.75$ \\
$2.0 \cdot 10^{8}$ & $-0.07$ & $0.15$ & $0.18$ & $0.43$ & $1.69$ \\
$3.0 \cdot 10^{8}$ & $0.10$ & $0.26$ & $0.23$ & $0.51$ & $1.82$ \\
$4.0 \cdot 10^{8}$ & $0.18$ & $0.34$ & $0.26$ & $0.56$ & $1.87$ \\
$5.0 \cdot 10^{8}$ & $0.23$ & $0.44$ & $0.30$ & $0.63$ & $1.99$ \\
$6.0 \cdot 10^{8}$ & $0.25$ & $0.51$ & $0.33$ & $0.68$ & $2.06$ \\
$7.0 \cdot 10^{8}$ & $0.27$ & $0.58$ & $0.36$ & $0.74$ & $2.16$ \\
$8.0 \cdot 10^{8}$ & $0.28$ & $0.64$ & $0.39$ & $0.79$ & $2.25$ \\
$9.0 \cdot 10^{8}$ & $0.29$ & $0.70$ & $0.42$ & $0.83$ & $2.33$ \\
$1.0 \cdot 10^{9}$ & $0.36$ & $0.79$ & $0.46$ & $0.91$ & $2.47$ \\
$1.0 \cdot 10^{9}$ & $0.36$ & $0.79$ & $0.46$ & $0.91$ & $2.47$ \\
$2.0 \cdot 10^{9}$ & $0.45$ & $0.89$ & $0.54$ & $1.11$ & $3.04$ \\
$3.0 \cdot 10^{9}$ & $0.53$ & $0.94$ & $0.57$ & $1.17$ & $3.21$ \\
$4.0 \cdot 10^{9}$ & $0.67$ & $1.02$ & $0.62$ & $1.26$ & $3.43$ \\
$5.0 \cdot 10^{9}$ & $0.69$ & $1.03$ & $0.62$ & $1.26$ & $3.41$ \\
$6.0 \cdot 10^{9}$ & $0.76$ & $1.06$ & $0.63$ & $1.29$ & $3.47$ \\
$7.0 \cdot 10^{9}$ & $0.81$ & $1.08$ & $0.65$ & $1.32$ & $3.53$ \\
$8.0 \cdot 10^{9}$ & $0.83$ & $1.08$ & $0.65$ & $1.32$ & $3.55$ \\
$9.0 \cdot 10^{9}$ & $0.87$ & $1.10$ & $0.66$ & $1.34$ & $3.58$ \\
$1.0 \cdot 10^{10}$ & $0.89$ & $1.11$ & $0.66$ & $1.36$ & $3.61$ \\
$1.1 \cdot 10^{10}$ & $0.91$ & $1.12$ & $0.67$ & $1.36$ & $3.66$ \\
$1.2 \cdot 10^{10}$ & $0.92$ & $1.12$ & $0.67$ & $1.36$ & $3.62$ \\
$1.3 \cdot 10^{10}$ & $0.95$ & $1.13$ & $0.67$ & $1.37$ & $3.63$ \\
$1.4 \cdot 10^{10}$ & $0.97$ & $1.14$ & $0.67$ & $1.38$ & $3.63$ \\
$1.5 \cdot 10^{10}$ & $0.98$ & $1.14$ & $0.68$ & $1.39$ & $3.67$ \\
$1.6 \cdot 10^{10}$ & $1.00$ & $1.15$ & $0.68$ & $1.39$ & $3.69$ \\
\hline
\end{tabular}
\end{table}

\begin{table*}
\caption{Indices. Time is in years. See Worthey (\cite{worthey}) for index definitions.}
\label{tab_idx1}
\begin{tabular}{r|r|r|r|r|r|r|r|r|r}
time&	H$_{\beta}$&	$Mg_1$&	Mg$_2$&	Mg$_b$&	Fe5270&	Fe5335&	NaD&	TiO$_1$&	TiO$_2$	\\
\hline
\hline
\multicolumn{10}{c}{Z = 0.0001}\\
\hline
$5.0 \cdot 10^{8}$ & $6.50$ & $0.020$ & $-0.003$ & $0.11$ & $-1.31$ & $0.20$ & $2.12$ & $0.013$ & $-0.006$ \\
$1.0 \cdot 10^{9}$ & $6.38$ & $0.014$ & $0.002$ & $0.30$ & $-0.99$ & $0.31$ & $2.41$ & $0.013$ & $-0.005$ \\
$2.0 \cdot 10^{9}$ & $4.94$ & $0.003$ & $0.009$ & $0.31$ & $-0.49$ & $0.38$ & $2.36$ & $0.013$ & $-0.003$ \\
$3.0 \cdot 10^{9}$ & $4.64$ & $0.001$ & $0.013$ & $0.44$ & $-0.36$ & $0.44$ & $2.44$ & $0.013$ & $-0.003$ \\
$4.0 \cdot 10^{9}$ & $4.21$ & $0.000$ & $0.015$ & $0.49$ & $-0.23$ & $0.48$ & $2.48$ & $0.013$ & $-0.003$ \\
$5.0 \cdot 10^{9}$ & $3.89$ & $-0.000$ & $0.017$ & $0.52$ & $-0.16$ & $0.51$ & $2.50$ & $0.012$ & $-0.003$ \\
$6.0 \cdot 10^{9}$ & $3.57$ & $0.000$ & $0.019$ & $0.53$ & $-0.09$ & $0.53$ & $2.50$ & $0.012$ & $-0.004$ \\
$7.0 \cdot 10^{9}$ & $3.33$ & $0.001$ & $0.019$ & $0.51$ & $-0.03$ & $0.56$ & $2.54$ & $0.012$ & $-0.004$ \\
$8.0 \cdot 10^{9}$ & $3.13$ & $0.001$ & $0.018$ & $0.47$ & $0.02$ & $0.58$ & $2.57$ & $0.012$ & $-0.004$ \\
$9.0 \cdot 10^{9}$ & $3.01$ & $0.002$ & $0.018$ & $0.43$ & $0.05$ & $0.61$ & $2.61$ & $0.011$ & $-0.005$ \\
$1.0 \cdot 10^{10}$ & $2.95$ & $0.003$ & $0.018$ & $0.44$ & $0.07$ & $0.62$ & $2.62$ & $0.011$ & $-0.005$ \\
$1.1 \cdot 10^{10}$ & $2.90$ & $0.004$ & $0.019$ & $0.46$ & $0.09$ & $0.63$ & $2.64$ & $0.011$ & $-0.005$ \\
$1.2 \cdot 10^{10}$ & $2.90$ & $0.004$ & $0.020$ & $0.48$ & $0.10$ & $0.64$ & $2.65$ & $0.011$ & $-0.005$ \\
$1.3 \cdot 10^{10}$ & $2.92$ & $0.005$ & $0.021$ & $0.50$ & $0.11$ & $0.65$ & $2.64$ & $0.011$ & $-0.006$ \\
$1.4 \cdot 10^{10}$ & $2.94$ & $0.007$ & $0.023$ & $0.53$ & $0.11$ & $0.66$ & $2.68$ & $0.010$ & $-0.006$ \\
$1.5 \cdot 10^{10}$ & $2.92$ & $0.008$ & $0.024$ & $0.54$ & $0.12$ & $0.67$ & $2.69$ & $0.010$ & $-0.006$ \\
$1.6 \cdot 10^{10}$ & $2.93$ & $0.010$ & $0.026$ & $0.57$ & $0.12$ & $0.69$ & $2.74$ & $0.010$ & $-0.007$ \\
\hline
\multicolumn{10}{c}{Z = 0.0004}\\
\hline
$5.0 \cdot 10^{8}$ & $6.14$ & $0.010$ & $0.014$ & $0.35$ & $-0.60$ & $0.02$ & $1.85$ & $0.013$ & $-0.005$ \\
$1.0 \cdot 10^{9}$ & $5.33$ & $0.004$ & $0.023$ & $0.35$ & $-0.18$ & $0.22$ & $2.15$ & $0.013$ & $-0.003$ \\
$2.0 \cdot 10^{9}$ & $4.43$ & $0.003$ & $0.026$ & $0.68$ & $0.30$ & $0.43$ & $2.05$ & $0.013$ & $-0.001$ \\
$3.0 \cdot 10^{9}$ & $4.16$ & $0.003$ & $0.028$ & $0.71$ & $0.37$ & $0.46$ & $2.04$ & $0.013$ & $-0.001$ \\
$4.0 \cdot 10^{9}$ & $3.79$ & $0.003$ & $0.031$ & $0.76$ & $0.47$ & $0.52$ & $2.08$ & $0.013$ & $-0.001$ \\
$5.0 \cdot 10^{9}$ & $3.47$ & $0.003$ & $0.033$ & $0.80$ & $0.54$ & $0.56$ & $2.10$ & $0.013$ & $-0.001$ \\
$6.0 \cdot 10^{9}$ & $3.17$ & $0.004$ & $0.038$ & $0.88$ & $0.62$ & $0.61$ & $2.14$ & $0.013$ & $-0.001$ \\
$7.0 \cdot 10^{9}$ & $2.97$ & $0.006$ & $0.042$ & $0.94$ & $0.67$ & $0.65$ & $2.18$ & $0.012$ & $-0.002$ \\
$8.0 \cdot 10^{9}$ & $2.80$ & $0.006$ & $0.043$ & $0.95$ & $0.72$ & $0.68$ & $2.21$ & $0.012$ & $-0.002$ \\
$9.0 \cdot 10^{9}$ & $2.66$ & $0.007$ & $0.044$ & $0.93$ & $0.76$ & $0.70$ & $2.23$ & $0.012$ & $-0.002$ \\
$1.0 \cdot 10^{10}$ & $2.58$ & $0.008$ & $0.044$ & $0.90$ & $0.78$ & $0.72$ & $2.28$ & $0.012$ & $-0.002$ \\
$1.1 \cdot 10^{10}$ & $2.50$ & $0.009$ & $0.044$ & $0.88$ & $0.80$ & $0.74$ & $2.31$ & $0.012$ & $-0.002$ \\
$1.2 \cdot 10^{10}$ & $2.42$ & $0.009$ & $0.044$ & $0.85$ & $0.83$ & $0.76$ & $2.31$ & $0.011$ & $-0.002$ \\
$1.3 \cdot 10^{10}$ & $2.38$ & $0.010$ & $0.045$ & $0.84$ & $0.85$ & $0.77$ & $2.32$ & $0.011$ & $-0.002$ \\
$1.4 \cdot 10^{10}$ & $2.42$ & $0.011$ & $0.046$ & $0.85$ & $0.85$ & $0.78$ & $2.32$ & $0.011$ & $-0.003$ \\
$1.5 \cdot 10^{10}$ & $2.59$ & $0.012$ & $0.048$ & $0.89$ & $0.81$ & $0.77$ & $2.33$ & $0.011$ & $-0.003$ \\
$1.6 \cdot 10^{10}$ & $2.58$ & $0.013$ & $0.049$ & $0.91$ & $0.83$ & $0.78$ & $2.33$ & $0.011$ & $-0.003$ \\
\hline
\multicolumn{10}{c}{Z = 0.004}\\
\hline
$5.0 \cdot 10^{8}$ & $6.10$ & $0.016$ & $0.048$ & $0.85$ & $0.56$ & $0.39$ & $1.50$ & $0.016$ & $0.003$ \\
$1.0 \cdot 10^{9}$ & $5.31$ & $0.017$ & $0.061$ & $1.14$ & $0.96$ & $0.70$ & $1.80$ & $0.014$ & $0.001$ \\
$2.0 \cdot 10^{9}$ & $3.66$ & $0.030$ & $0.095$ & $1.64$ & $1.53$ & $1.15$ & $1.72$ & $0.020$ & $0.016$ \\
$3.0 \cdot 10^{9}$ & $2.79$ & $0.037$ & $0.111$ & $1.86$ & $1.78$ & $1.35$ & $1.72$ & $0.024$ & $0.024$ \\
$4.0 \cdot 10^{9}$ & $2.63$ & $0.038$ & $0.112$ & $1.84$ & $1.80$ & $1.38$ & $1.81$ & $0.025$ & $0.026$ \\
$5.0 \cdot 10^{9}$ & $2.42$ & $0.041$ & $0.117$ & $1.92$ & $1.86$ & $1.45$ & $1.91$ & $0.026$ & $0.029$ \\
$6.0 \cdot 10^{9}$ & $2.26$ & $0.044$ & $0.124$ & $2.02$ & $1.92$ & $1.51$ & $1.99$ & $0.028$ & $0.032$ \\
$7.0 \cdot 10^{9}$ & $2.12$ & $0.047$ & $0.129$ & $2.10$ & $1.98$ & $1.56$ & $2.04$ & $0.029$ & $0.034$ \\
$8.0 \cdot 10^{9}$ & $2.03$ & $0.049$ & $0.133$ & $2.18$ & $2.01$ & $1.60$ & $2.11$ & $0.029$ & $0.034$ \\
$9.0 \cdot 10^{9}$ & $1.97$ & $0.052$ & $0.138$ & $2.26$ & $2.04$ & $1.64$ & $2.20$ & $0.030$ & $0.036$ \\
$1.0 \cdot 10^{10}$ & $1.90$ & $0.054$ & $0.141$ & $2.32$ & $2.08$ & $1.68$ & $2.29$ & $0.030$ & $0.036$ \\
$1.1 \cdot 10^{10}$ & $1.86$ & $0.057$ & $0.146$ & $2.39$ & $2.11$ & $1.72$ & $2.38$ & $0.030$ & $0.037$ \\
$1.2 \cdot 10^{10}$ & $1.82$ & $0.059$ & $0.150$ & $2.45$ & $2.13$ & $1.76$ & $2.48$ & $0.031$ & $0.039$ \\
$1.3 \cdot 10^{10}$ & $1.78$ & $0.062$ & $0.155$ & $2.52$ & $2.16$ & $1.80$ & $2.57$ & $0.033$ & $0.043$ \\
$1.4 \cdot 10^{10}$ & $1.77$ & $0.062$ & $0.156$ & $2.54$ & $2.16$ & $1.80$ & $2.59$ & $0.032$ & $0.041$ \\
$1.5 \cdot 10^{10}$ & $1.77$ & $0.062$ & $0.158$ & $2.55$ & $2.16$ & $1.80$ & $2.60$ & $0.033$ & $0.042$ \\
$1.6 \cdot 10^{10}$ & $1.76$ & $0.063$ & $0.162$ & $2.57$ & $2.17$ & $1.82$ & $2.61$ & $0.033$ & $0.042$ \\
\hline
\end{tabular}
\end{table*}

\begin{table*}
\caption{continuation of table \ref{tab_idx1}.}
\label{tab_idx2}
\begin{tabular}{r|r|r|r|r|r|r|r|r|r}
time&	H$_{\beta}$&	$Mg_1$&	Mg$_2$&	Mg$_b$&	Fe5270&	Fe5335&	NaD&	TiO$_1$&	TiO$_2$	\\
\hline
\multicolumn{10}{c}{Z = 0.008}\\
\hline
$5.0 \cdot 10^{8}$ & $6.41$ & $0.018$ & $0.057$ & $0.98$ & $0.84$ & $0.59$ & $1.39$ & $0.016$ & $0.004$ \\
$1.0 \cdot 10^{9}$ & $5.05$ & $0.022$ & $0.076$ & $1.38$ & $1.31$ & $1.00$ & $1.69$ & $0.016$ & $0.005$ \\
$2.0 \cdot 10^{9}$ & $3.16$ & $0.044$ & $0.128$ & $2.01$ & $1.95$ & $1.57$ & $1.94$ & $0.030$ & $0.035$ \\
$3.0 \cdot 10^{9}$ & $2.50$ & $0.055$ & $0.147$ & $2.32$ & $2.19$ & $1.81$ & $2.13$ & $0.034$ & $0.044$ \\
$4.0 \cdot 10^{9}$ & $2.41$ & $0.055$ & $0.149$ & $2.38$ & $2.17$ & $1.81$ & $2.23$ & $0.034$ & $0.043$ \\
$5.0 \cdot 10^{9}$ & $2.25$ & $0.058$ & $0.157$ & $2.52$ & $2.23$ & $1.88$ & $2.32$ & $0.035$ & $0.046$ \\
$6.0 \cdot 10^{9}$ & $2.05$ & $0.065$ & $0.170$ & $2.68$ & $2.34$ & $1.99$ & $2.41$ & $0.038$ & $0.051$ \\
$7.0 \cdot 10^{9}$ & $1.97$ & $0.067$ & $0.175$ & $2.76$ & $2.38$ & $2.03$ & $2.48$ & $0.038$ & $0.052$ \\
$8.0 \cdot 10^{9}$ & $1.89$ & $0.070$ & $0.180$ & $2.84$ & $2.42$ & $2.08$ & $2.55$ & $0.039$ & $0.053$ \\
$9.0 \cdot 10^{9}$ & $1.82$ & $0.072$ & $0.184$ & $2.90$ & $2.45$ & $2.11$ & $2.60$ & $0.038$ & $0.052$ \\
$1.0 \cdot 10^{10}$ & $1.76$ & $0.075$ & $0.189$ & $2.96$ & $2.49$ & $2.15$ & $2.66$ & $0.039$ & $0.054$ \\
$1.1 \cdot 10^{10}$ & $1.70$ & $0.077$ & $0.194$ & $3.02$ & $2.53$ & $2.19$ & $2.71$ & $0.040$ & $0.055$ \\
$1.2 \cdot 10^{10}$ & $1.64$ & $0.080$ & $0.198$ & $3.07$ & $2.56$ & $2.23$ & $2.77$ & $0.041$ & $0.057$ \\
$1.3 \cdot 10^{10}$ & $1.59$ & $0.082$ & $0.200$ & $3.09$ & $2.58$ & $2.25$ & $2.80$ & $0.039$ & $0.054$ \\
$1.4 \cdot 10^{10}$ & $1.56$ & $0.084$ & $0.203$ & $3.13$ & $2.60$ & $2.27$ & $2.86$ & $0.042$ & $0.059$ \\
$1.5 \cdot 10^{10}$ & $1.54$ & $0.085$ & $0.206$ & $3.16$ & $2.61$ & $2.29$ & $2.93$ & $0.042$ & $0.059$ \\
$1.6 \cdot 10^{10}$ & $1.52$ & $0.087$ & $0.208$ & $3.19$ & $2.62$ & $2.32$ & $2.99$ & $0.042$ & $0.059$ \\
\hline
\multicolumn{10}{c}{Z = 0.02}\\
\hline
$5.0 \cdot 10^{8}$ & $6.41$ & $0.018$ & $0.079$ & $1.22$ & $1.28$ & $1.00$ & $1.33$ & $0.020$ & $0.010$ \\
$1.0 \cdot 10^{9}$ & $4.44$ & $0.027$ & $0.108$ & $1.78$ & $1.80$ & $1.51$ & $1.70$ & $0.021$ & $0.014$ \\
$2.0 \cdot 10^{9}$ & $2.78$ & $0.066$ & $0.172$ & $2.58$ & $2.43$ & $2.18$ & $2.48$ & $0.036$ & $0.049$ \\
$3.0 \cdot 10^{9}$ & $2.20$ & $0.084$ & $0.204$ & $3.10$ & $2.72$ & $2.47$ & $2.80$ & $0.039$ & $0.056$ \\
$4.0 \cdot 10^{9}$ & $2.02$ & $0.090$ & $0.216$ & $3.28$ & $2.80$ & $2.56$ & $2.93$ & $0.040$ & $0.058$ \\
$5.0 \cdot 10^{9}$ & $1.90$ & $0.095$ & $0.225$ & $3.41$ & $2.86$ & $2.63$ & $3.04$ & $0.040$ & $0.059$ \\
$6.0 \cdot 10^{9}$ & $1.81$ & $0.100$ & $0.233$ & $3.51$ & $2.91$ & $2.69$ & $3.15$ & $0.042$ & $0.062$ \\
$7.0 \cdot 10^{9}$ & $1.74$ & $0.102$ & $0.237$ & $3.58$ & $2.94$ & $2.73$ & $3.23$ & $0.042$ & $0.062$ \\
$8.0 \cdot 10^{9}$ & $1.68$ & $0.104$ & $0.242$ & $3.65$ & $2.98$ & $2.77$ & $3.30$ & $0.042$ & $0.063$ \\
$9.0 \cdot 10^{9}$ & $1.60$ & $0.109$ & $0.250$ & $3.73$ & $3.04$ & $2.84$ & $3.39$ & $0.043$ & $0.065$ \\
$1.0 \cdot 10^{10}$ & $1.54$ & $0.113$ & $0.256$ & $3.79$ & $3.08$ & $2.89$ & $3.48$ & $0.044$ & $0.066$ \\
$1.1 \cdot 10^{10}$ & $1.49$ & $0.116$ & $0.261$ & $3.85$ & $3.12$ & $2.93$ & $3.55$ & $0.045$ & $0.069$ \\
$1.2 \cdot 10^{10}$ & $1.44$ & $0.118$ & $0.265$ & $3.89$ & $3.15$ & $2.96$ & $3.61$ & $0.045$ & $0.069$ \\
$1.3 \cdot 10^{10}$ & $1.39$ & $0.121$ & $0.269$ & $3.93$ & $3.18$ & $3.00$ & $3.67$ & $0.045$ & $0.069$ \\
$1.4 \cdot 10^{10}$ & $1.35$ & $0.123$ & $0.273$ & $3.97$ & $3.21$ & $3.03$ & $3.73$ & $0.046$ & $0.072$ \\
$1.5 \cdot 10^{10}$ & $1.32$ & $0.125$ & $0.276$ & $4.01$ & $3.23$ & $3.06$ & $3.78$ & $0.046$ & $0.072$ \\
$1.6 \cdot 10^{10}$ & $1.29$ & $0.127$ & $0.278$ & $4.03$ & $3.25$ & $3.08$ & $3.83$ & $0.047$ & $0.073$ \\
\hline
\multicolumn{10}{c}{Z = 0.05}\\
\hline
$5.0 \cdot 10^{8}$ & $5.79$ & $0.035$ & $0.116$ & $1.70$ & $1.89$ & $1.74$ & $1.87$ & $0.023$ & $0.021$ \\
$1.0 \cdot 10^{9}$ & $3.15$ & $0.079$ & $0.194$ & $2.84$ & $2.78$ & $2.66$ & $2.89$ & $0.026$ & $0.032$ \\
$2.0 \cdot 10^{9}$ & $2.29$ & $0.103$ & $0.235$ & $3.46$ & $3.04$ & $3.01$ & $3.29$ & $0.041$ & $0.065$ \\
$3.0 \cdot 10^{9}$ & $1.97$ & $0.114$ & $0.256$ & $3.81$ & $3.18$ & $3.18$ & $3.52$ & $0.043$ & $0.070$ \\
$4.0 \cdot 10^{9}$ & $1.66$ & $0.140$ & $0.294$ & $4.22$ & $3.45$ & $3.48$ & $4.01$ & $0.049$ & $0.084$ \\
$5.0 \cdot 10^{9}$ & $1.59$ & $0.140$ & $0.297$ & $4.30$ & $3.46$ & $3.50$ & $4.06$ & $0.048$ & $0.082$ \\
$6.0 \cdot 10^{9}$ & $1.46$ & $0.149$ & $0.311$ & $4.44$ & $3.55$ & $3.61$ & $4.25$ & $0.050$ & $0.086$ \\
$7.0 \cdot 10^{9}$ & $1.38$ & $0.156$ & $0.322$ & $4.54$ & $3.62$ & $3.69$ & $4.41$ & $0.052$ & $0.090$ \\
$8.0 \cdot 10^{9}$ & $1.33$ & $0.159$ & $0.327$ & $4.60$ & $3.66$ & $3.73$ & $4.52$ & $0.052$ & $0.090$ \\
$9.0 \cdot 10^{9}$ & $1.27$ & $0.164$ & $0.335$ & $4.68$ & $3.71$ & $3.79$ & $4.66$ & $0.053$ & $0.093$ \\
$1.0 \cdot 10^{10}$ & $1.23$ & $0.168$ & $0.342$ & $4.74$ & $3.76$ & $3.84$ & $4.79$ & $0.054$ & $0.096$ \\
$1.1 \cdot 10^{10}$ & $1.20$ & $0.170$ & $0.345$ & $4.77$ & $3.79$ & $3.88$ & $4.89$ & $0.055$ & $0.096$ \\
$1.2 \cdot 10^{10}$ & $1.17$ & $0.172$ & $0.349$ & $4.81$ & $3.82$ & $3.91$ & $5.00$ & $0.055$ & $0.095$ \\
$1.3 \cdot 10^{10}$ & $1.13$ & $0.175$ & $0.355$ & $4.86$ & $3.86$ & $3.95$ & $5.11$ & $0.055$ & $0.096$ \\
$1.4 \cdot 10^{10}$ & $1.10$ & $0.178$ & $0.359$ & $4.89$ & $3.89$ & $3.99$ & $5.20$ & $0.055$ & $0.096$ \\
$1.5 \cdot 10^{10}$ & $1.07$ & $0.180$ & $0.362$ & $4.92$ & $3.91$ & $4.02$ & $5.28$ & $0.056$ & $0.098$ \\
$1.6 \cdot 10^{10}$ & $1.05$ & $0.182$ & $0.365$ & $4.94$ & $3.93$ & $4.04$ & $5.34$ & $0.056$ & $0.098$ \\
\hline
\end{tabular}
\end{table*}

\end{document}